# Mathematical Analysis of N-R Algorithm for Experimental Implementation of SHEPWM Control on Single-phase Inverter

Alla Eddine TOUBAL MAAMAR[#1], M'hamed HELAIMI[#2], Rachid TALEB[#3], Hadj MOULOUDJ[*4], Oumaymah ELAMRI[**5], Abdelatif GADOUM[#6]

[#]Electrical Engineering Department, Laboratoire Génie Electrique et Energies Renouvelables (LGEER), Hassiba Benbouali University of Chlef, Chlef, Algeria

[*]Faculté des Science Exactes et Informatique, Hassiba Benbouali University of Chlef
LPPPS Laboratory,  Ecole Normale Supérieure, Vieux Kouba, Alger, Algeria

[**]Laboratoire de Génie Electrique et Commande des Systèmes (LGECOS), Ecole Nationale des Sciences Appliquées de Marrakech,Cadi Ayyad University, Morocco

***Abstract*** — This paper presents a simulation and experimental validation of H-bridge inverter with the implementation of a Newton-Raphson algorithm for selective harmonics elimination. Contributions are made in interfacing and programming of Arduino ATmega328P microcontroller and use it to control of single-phase inverter by selective harmonic elimination technique. Through this work, we have given a general analytical description of the total harmonic distortion and present the method of application of the Newton-Raphson algorithm for selective harmonic elimination. MATLAB Simulink used for simulation and Arduino IDE software was used to program the Arduino board. The validity of the simulation model is verified by experimental results, the inverter was realized and the algorithm control is tested, results of the simulation and realization are compared, they show the efficiency of the system analysis and demonstrate the ability of Arduino to control the inverter and to generate complex signals for the electrical systems control.

## I. INTRODUCTION

Nowadays, alternative electrical energy takes a huge part of our lives, most of our electrical materials, electronic devices, industrial equipment are needs this energy form, for that the electrical signal quality obtained is a very important point in the conversion and distribution of electrical energy, one of the parameters that can provide this quality is the abbreviated term THD, Total Harmonic Distortion.

Over the last few years, and more exactly following to the transistor invention with the great development in the field of switching electronics and the removal of electrical transformers in some new electrical devices a new type of problem on the electrical networks has become apparent as a consequence of the increasing use of nonlinear electronic equipment connected to this electrical network. The nonlinear devices can affect the shape of the electrical network voltage by calling disruptive currents, this modification or distortion of the shape of the electrical voltage called harmonic pollution, among the most important nonlinear devices that generate harmonics, we can mention: induction heating furnaces, computers, discharge lamps, battery chargers, variable speed drives, frequency converters, inverters, electronic power supplies. The existence of harmonics in a circuit disrupts other loads with the generation of over-currents in the circuit or over-heating of equipment and the third harmonic plus their multiple are the cause of the neutral cables overheating [1], [2], [3].

Through this work we have given a general analytical description about the total harmonic distortion then present the method of application of the Newton-Raphson algorithm for the selective harmonic elimination, this method is proposed to eliminate the unwanted harmonics, the third harmonics is the most unwanted.

## II. TOTAL HARMONICS DISTORTION, RATINGS AND DEFINITIONS

Harmonics are sine wave voltages or currents whose frequency is an integer multiple of the network frequency when added to the fundamental sine wave voltage or current result the distortion of the energy waveform. If the THD is zero, we can say that there aren't harmonics in the electrical network. This total harmonic distortion corresponds to the ratio between the real (RMS) value of the harmonics of a signal (voltage or current) and its (RMS) value at the fundamental frequency. Two harmonic distortion rates are distinguished: The total harmonic distortion of voltage THDv, and The total harmonic distortion of current THDi [2], [3], [4]:

$$THD_v = \sqrt{\sum_{h=2}^{+\infty}\left(\frac{V_h}{V_1}\right)^2}, \quad THD_i = \sqrt{\sum_{h=2}^{+\infty}\left(\frac{I_h}{I_1}\right)^2}$$

Vh, Ih: The root means square (RMS) value of harmonic voltage and harmonic current.

V1, I1: The root means square (RMS) value of the fundamental voltage and the fundamental current.

The main harmonic generators are the non-linear loads that generate harmonics which flow into the electrical network from the load to the power supply. These harmonics are present in the form of " lines " called spectrum, such as harmonics over the 30th rank are neglected, even rank harmonics (2, 4, 6, 8) are almost zero due to the symmetry of the energy signal so they are also negligible and odd rank harmonics (3, 5, 7, 9...) are frequently observed in the electrical network. The existence of harmonics in an electrical network results to: the perturbation of other loads, effective current values less or higher than those required for the load's energy needs, they generate disturbances in the operation of protection equipment and switching devices, heating of the neutral cable: harmonic frequency currents of the third rank and our multiples are summed in the neutral conductor, the over-current in the electrical circuit, heating of electrical equipment and after some operation periods their failures[1], [5].

Selective Harmonic Elimination (SHE) is a together of nonlinear equations that can be solved to obtain the switching angles for the inverter [6]. SHE is proposed to eliminate the unwanted harmonics, the calculation of switching angles is very complicated for that many algorithms have been proposed but the most popular algorithms are: Newton Raphson algorithm (NR), Practical Swarm Optimization (PSO), Genetic Algorithm (GA), Artificial Neural Network (ANN).

## III. PROBLEM FORMULATION WITH MATHEMATICAL ASPECT

The proposed inverter for this study is the H-bridge inverter, the most used inverter then other common topologies (neutral-point clamped inverter (NPC), flying capacitor (FC) inverter) for their high efficiency and advantages [7].

The topology of H-bridge inverter is shown in Fig.1, consists of four electronic switches, to avoid voltage source short-circuit and to have a continuous conducting, an additional control must be defined.

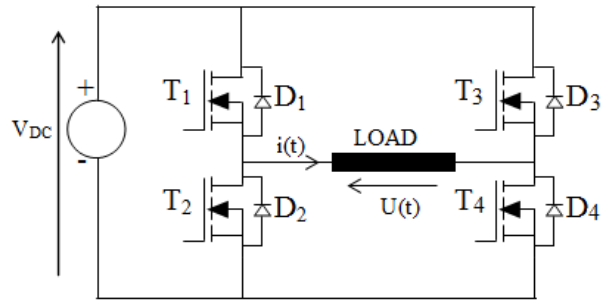

Fig. 1 The topology of the cascaded H-bridge Inverter

For inverter control and elimination of unwanted harmonics, a waveform similar to the generalized SHEPWM waveform shown in Fig.2 must be generated.

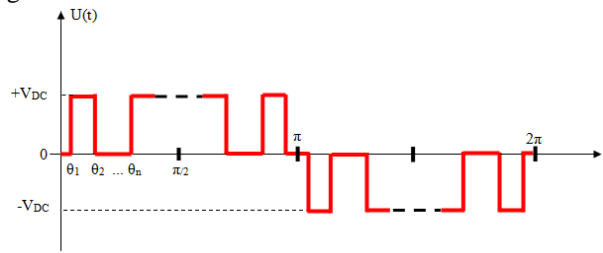

Fig. 2 A generalized SHE PWM waveform

The French mathematician Joseph Fourier (1768 - 1830) has demonstrated that any complex signal with conditions can be put into the sum of sinusoidal functions. Among the essential conditions, it is necessary that the signal is periodic and continue by parts. With the decomposition of a signal, the calculation becomes easy because we have processed separately the sinusoidal functions obtained (harmonics) and after we apply the superposition principle [8], [9], [10].

The Fourier series of a signal U(t) expressed by the mathematical expressions of the following trigonometric shape.

$$U(t) = a_0 + \sum_{n=1}^{\infty}(a_n \cos(nwt) + b_n \sin(nwt)) \quad (1)$$

n: Whole Numbers: 0, 1, 2, …n, ω=1/T , $\theta=\omega t$, ω: The pulsation

The Fourier coefficients ($a_0$, $a_n$, $b_n$) are given by the following integrals:

$$a_0 = \frac{1}{T}\int_0^T U(t)\,dt \quad (2)$$
$$a_n = \frac{2}{T}\int_0^T U(t)\cos(nwt)\,dt \quad (3)$$
$$b_n = \frac{2}{T}\int_0^T U(t)\sin(nwt)\,dt \quad (4)$$

A periodic function U(t) is odd, $a_0$=0, $a_n$=0, we calculate only the coefficient $b_n$.

$$b_n = \frac{4V}{n\pi}\sum_{i=1}^{p}(-1)^{i+1}\cos(n\theta_i) \quad (5)$$

p: The number of switching angles in the quarter waveform, n: The odd harmonic ranks.

For example, the number of switching angles p=2, the fundamental Harmonique h1 is given where (n=1) by:

$$h_1 = \frac{4V}{\pi}\sum_{i=1}^{2}(-1)^{i+1}\cos(\theta_i) = \frac{4V}{\pi}(\cos(\theta_1) - \cos(\theta_2))$$
(6)

The algebraic system with a non-linear equation for finding the switching angles is:

$$(\cos(\theta_1) - \cos(\theta_2)) = \frac{\pi h_1}{4V} = \frac{\pi}{4}M \quad (1)$$

M: Modulation Index is the relation between the fundamental voltage and maximum voltage, M= h1/V

The third Harmonique h3 is given where (n=3) by:

$$h_3 = \frac{4V}{3\pi}\sum_{i=1}^{2}(-1)^{i+1}\cos(3\theta_i)$$

$$h_3 = \frac{4V}{3\pi}(\cos(3\theta_1) - \cos(3\theta_2))\quad(2)$$

The Elimination of the third harmonic and their multiple is possible where:

$$(\cos(3\theta_1) - \cos(3\theta_2)) = 0 \quad (3)$$

## IV. NEWTON-RAPHSON ALGORITHM

We have an algebraic system with non-linear equations (8 and 9) for finding the switching angles. The algorithm of the newton-raphson method presented can be used to solve such a system [11]. For example, if we want to eliminate the third harmonic and their multiple, we must search the switching angles θ1, θ2.

The algebraic system of nonlinear equations:

$$\cos(\theta_1) - \cos(\theta_2) = \frac{\pi}{4}M \quad (4)$$
$$\cos(3\theta_1) - \cos(3\theta_2) = 0 \quad (5)$$

The initial values of the switching angle:
$$\theta^j = [\theta_1^j;\ \theta_2^j]$$

The nonlinear system can be written: $F(\theta^j) = T$

$$F(\theta^j) = \begin{bmatrix} \cos(\theta_1^j) & -\cos(\theta_2^j) \\ \cos(3\theta_1^j) & -\cos(3\theta_2^j) \end{bmatrix}$$

$$T = \begin{bmatrix} M\frac{\pi}{4} & 0 \end{bmatrix}^T$$

Derivate of the nonlinear system matrix: $F(\theta^j)$

$$\left[\frac{\partial F(\theta)}{\partial \theta}\right]^j = \begin{bmatrix} -\sin(\theta_1^j) & \sin(\theta_2^j) \\ -3\sin(3\theta_1^j) & 3\sin(3\theta_2^j) \end{bmatrix}$$

The statement of the N-R algorithm is:
1. Set of initial values for $\theta^j$ with j=0

$$\theta^0 = [\theta_1^0 \quad \theta_2^0]^T$$

2. Calculate the value of: $F(\theta^0) = F^0$
3. Linearize of system equations about $\theta^0$

$$F^0 + \left[\frac{\partial F}{\partial \theta}\right]^0 d\theta^0 = T \quad (6)$$

Such as:
$$d\theta^0 = [d\theta_1^0 \quad d\theta_2^0]^T$$

4. Solve $d\theta^0$ from equation:

$$d\theta^0 = \left(inv\left(\left[\frac{\partial F}{\partial \theta}\right]^0\right)\right)(T - F^0) \quad (7)$$

5. Update the initial values: $\theta^{j+1} = \theta^j + d\theta^j$
6. Repeat the process for equations until:
$d\theta^j$ is satisfied with the desired degree of accuracy
The solution must satisfy the condition:

$$0 < \theta_1 < \theta_2 < \frac{\pi}{2}$$

## V. SIMULATION RESULTS

In this section, MATLAB/Simulation results are presented.

### A. Third Harmonic Elimination with their Multiple

To eliminate the third harmonic with their multiple need to calculate two switching angles, θ1 and θ1, after programming and executing of the N-R algorithm in the Matlab platform we have obtained the following results: With modulation index, M= 0.85, switching angles are: θ1 =37.33°, θ2 =82.67°.

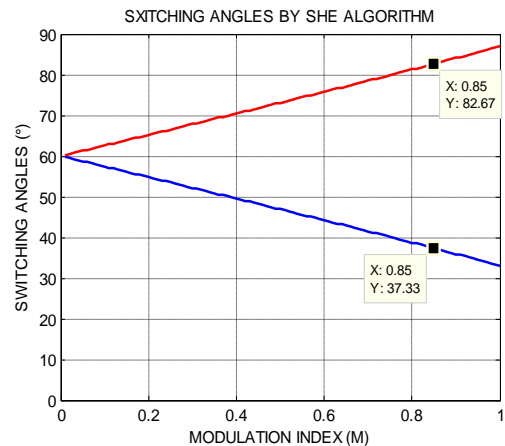

Fig. 3 Switching angles by Newton-Raphson algorithm with M=0.85

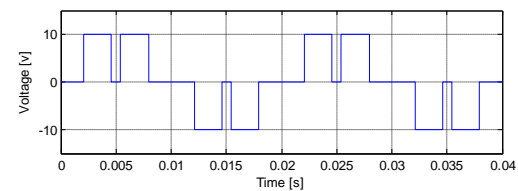

Fig. 4 The output voltage waveform of H-bridge inverter

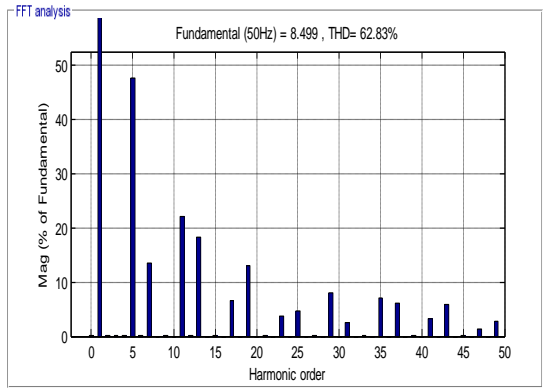

Fig. 5 The harmonic spectrum of the output voltage waveform

### B. Elimination of Third and fifth Harmonics

To eliminate the third harmonic and fifth harmonic need to calculate three switching angles, after programming and executing of the N-R algorithm in the Matlab platform we have obtained the following results:

With modulation index M= 0.85, switching angles are: θ1 =30.45°, θ2 =54.28°, θ3 =67.09°.

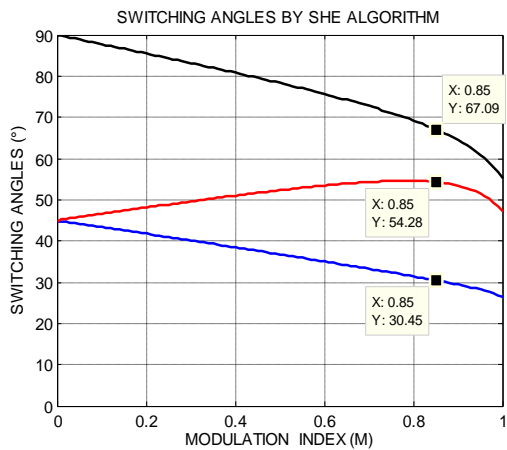

Fig. 6 Switching angles by Newton-Raphson algorithm with M=0.85

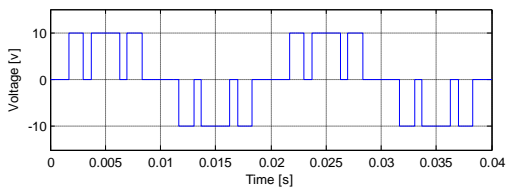

Fig. 7 The output voltage waveform of H-bridge inverter

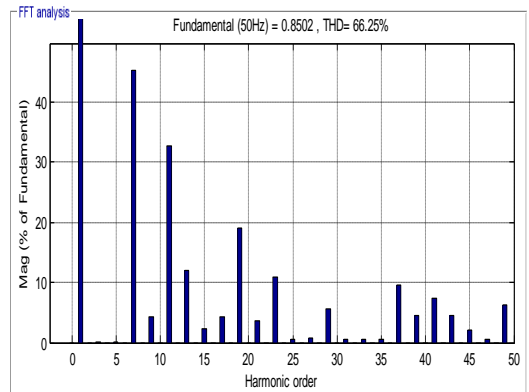

Fig. 8 The harmonic spectrum of the output voltage waveform

### C. Eliminations of four first Odd Harmonics

To eliminate the four first odd harmonics need to calculate five switching angles, after programming and executing of the N-R algorithm in the Matlab platform we have obtained the following results:

With modulation index M= 0.85, switching angles are: θ1=22.58°, θ2=33.6°, θ3=46.64°, θ4=68.5°, θ5=75.1°

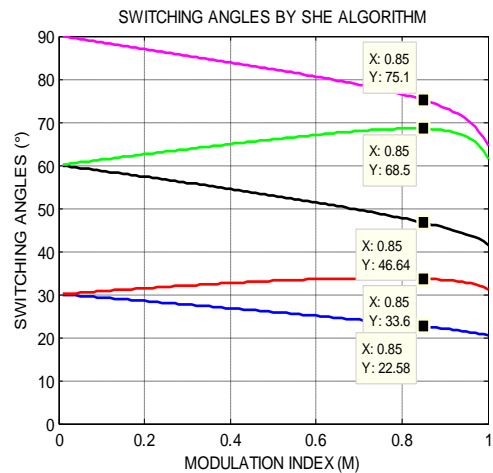

Fig. 9 Switching angles by Newton-Raphson algorithm with M=0.85

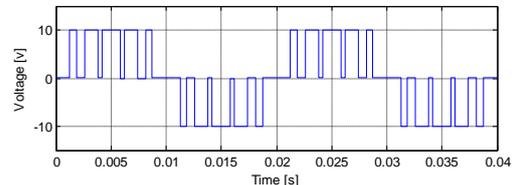

Fig. 10 The output voltage waveform of H-bridge inverter

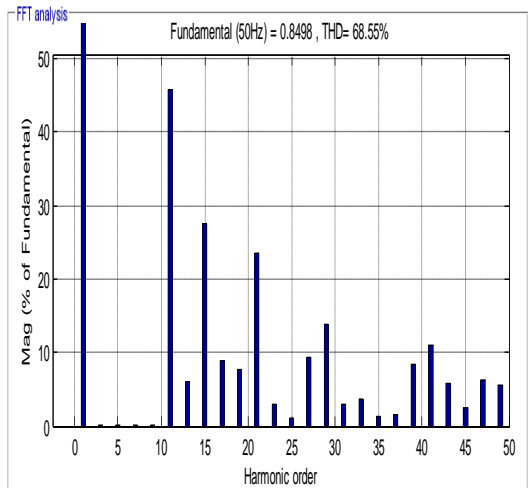

Fig. 11The harmonic spectrum of the output voltage waveform

The Newton-Raphson algorithm was programmed on the Matlab environment to determine the angles (Fig. 3), (Fig. 6), (Fig. 9) that will eliminate some unwanted harmonics.

The figures (Fig. 4) (Fig. 7) (Fig. 10): are the output voltage waveforms of the inverter with selective harmonic elimination control. In figure (Fig. 5) of the harmonic spectrum, it is clearly identified that the $3^{rd}$ harmonics and their multiple are completely eliminated, in figure (Fig. 8) of the harmonic spectrum, the $3^{rd}$ and $5^{th}$ harmonics are completely eliminated and in figure (Fig. 11) of the harmonic spectrum, the $3^{th}$, $5^{th}$, $7^{th}$, $9^{th}$ harmonics are completely eliminated. We can say that the method gives a good result because the selected harmonics are really eliminated, the THD can be decrease with optimum methods [12].

## VI. EXPERIMENTAL RESULTS

A first step is programming ARDUINO Uno ATmega 320P microcontroller with ARDUINO IDE software for generation of inverter control signals and next step is the implementation of signals in the experimental prototype, the figure (Figure 12) shows the control signals generated by Arduino for elimination of the $3^{rd}$ harmonics and their multiple, the figure (Figure 13) shows the control signals generated by Arduino for elimination of the $3^{rd}$ and $5^{th}$ harmonics, and the figure (Figure 14) shows the control signals generated by Arduino for elimination of the $3^{th}$, $5^{th}$, $7^{th}$, $9^{th}$ harmonics.

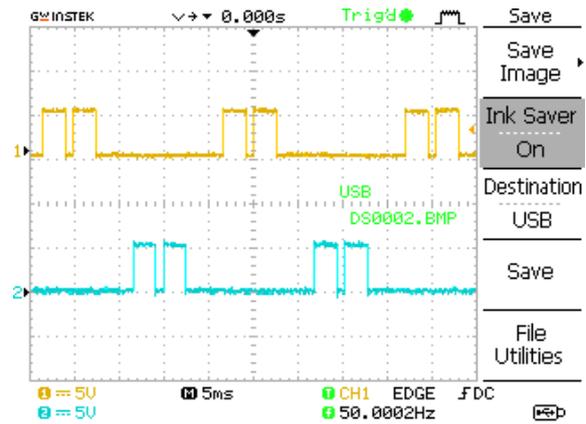

Fig. 12The control signals generated by Arduino

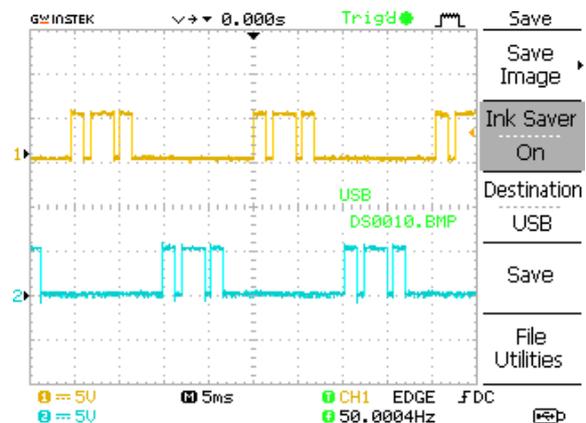

Fig. 13The control signals generated by Arduino

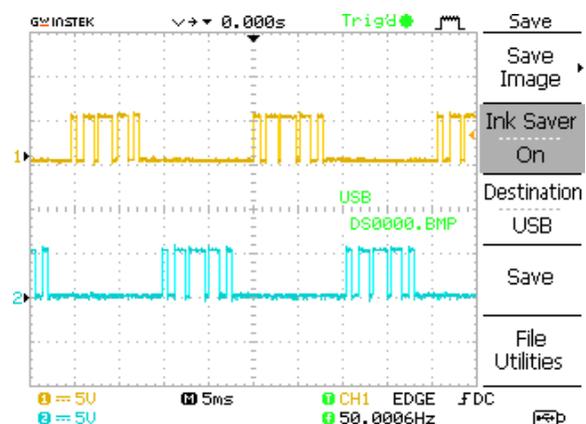

Fig. 14The control signals generated by Arduino

The prototype of the experimental system shown in Figure 15, consists of DC power supply, PC with ARDUINO IDE software, and cascaded H-bridge consist of four MOSFET switches IRF 640 controlled by driver circuits with TLP 250 and ARDUINO Uno ATmega 328P Board.

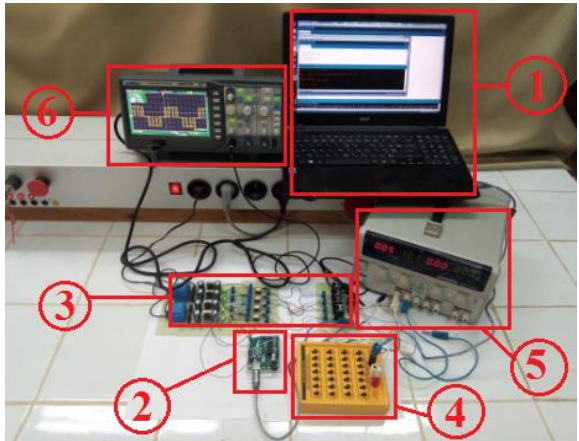

Fig. 15 The Experimental prototype of a system

1: Pc + ARDUINO IDE software.
2: Arduino Uno ATmega 328P Board.
3: H-bridge inverter + Driver circuits.
4: Resistor Load.
5: DC power supply.
6: Digital Oscilloscope.

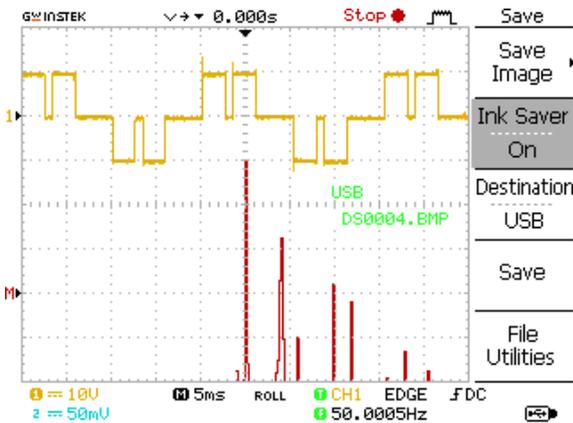

Fig. 16 The experimental output voltage waveform of H-bridge inverter

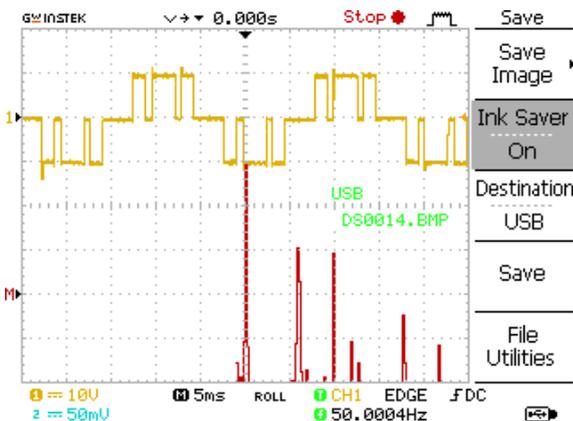

Fig. 17 The experimental output voltage waveform of H-bridge inverter

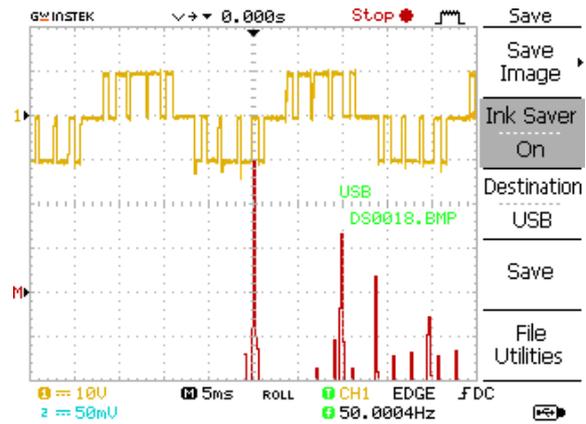

Fig. 18 The experimental output voltage waveform of H-bridge inverter

The figures (Fig. 16) (Fig. 17) (Fig. 18): shows the experimental output voltage waveforms of the inverter and harmonic spectrum. From figure (Fig. 16), the $3^{rd}$ harmonics and their multiple are eliminated, in figure (Fig. 17) the $3^{rd}$ and $5^{th}$ harmonics are completely eliminated and in figure (Fig. 18), the fourth odd harmonics ($3^{rd}$, $5^{th}$, $7^{th}$, $9^{th}$) are completely eliminated. The obtained results show the good concordance existing between the simulation and the realization system.

## VII. CONCLUSIONS

The simulation, realization and selective harmonic elimination control of a single-phase cascaded H-bridge inverter are presented in this paper, we have talked about the total harmonic distortion (THD) and presented the implementation of selective harmonic elimination (SHE) technique in H-bridge inverter, the mentioned objectives are elimination of the $3^{rd}$ harmonic and their multiple, then elimination of other rank harmonics unwanted in network using the Newton-Raphson algorithm, the obtained results (simulation and experimental) demonstrate the method quality.

The harmonics can't be removed but can reduce them according to the international electricity standards. The selective harmonic elimination method is generally based on ideas of opposite harmonic injection. Selective harmonic elimination (SHE) is a technique to eliminate harmonic ranges unwanted in the network with any influence on THD decrease, if you want to reduce the THD you should associate SHE with other optimization techniques or increase the level of inverter, use a filtering system or isolation transformers, implementation of phase-shifted control of the electronic switches, or with modified PWM using Genetic Algorithm. This work opens new ways for future research with other topologies (increase the level of the inverter) or other control cards, FPGA, DSP.

## APPENDIX A

The general algorithm of a Newton-Raphson algorithm for selective harmonic elimination

When the user needs to eliminate other selected harmonics, he can follow the following steps, the algebraic system of nonlinear equations:

$$\begin{cases} \cos(\theta_1) - \cos(\theta_2) + \cdots \mp \cos(\theta_P) = M\frac{\pi}{4} \\ \vdots \\ \cos(n\theta_1) - \cos(n\theta_2) + \cdots \mp \cos(n\theta_P) = 0 \end{cases}$$

n: The odd harmonic ranks
P: The number of switching angles in a quarter waveform.
M: Modulation Index, where M=h1/V
The switching angles:

$$\theta^j = [\theta_1^j; \theta_2^j; \ldots; \theta_P^j]$$

The nonlinear system can be written:

$$F(\theta^j) = T$$

$$F(\theta^j) = \begin{pmatrix} \cos(\theta_1^j) & -\cos(\theta_2^j) & \cdots & \mp \cos(\theta_P^j) \\ \vdots & \vdots & \ddots & \vdots \\ \cos(n\theta_1^j) & -\cos(n\theta_2^j) & \cdots & \mp \cos(n\theta_P^j) \end{pmatrix}$$

$$T = \begin{pmatrix} M\frac{\pi}{4} \\ \vdots \\ 0 \end{pmatrix}$$

Derivate of the nonlinear system matrix:

$$\left[\frac{\partial F(\theta)}{\partial \theta}\right]^j = \begin{pmatrix} -\sin(\theta_1^j) & \sin(\theta_2^j) & \cdots & \mp \sin(\theta_P^j) \\ \vdots & \vdots & \ddots & \vdots \\ -n\sin(n\theta_1^j) & n\sin(n\theta_2^j) & \cdots & \mp n\sin(n\theta_P^j) \end{pmatrix}$$

The statement of the algorithm:
1. Set of initial values for $\theta^j$ with j=0:

$$\theta^j = [\theta_1^j; \theta_2^j; \ldots; \theta_P^j]$$

2. Calculate the value of: $F(\theta^j) = F^j$

3. Linearize of the system equations about $\theta^j$

$$F^j + \left[\frac{\partial F}{\partial \theta}\right]^j d\theta^j = T$$

Such as: $d\theta^j = [d\theta_1^j \cdots d\theta_P^j]^T$

4. Solve $d\theta^j$ from equation:

$$d\theta^j = \left(inv\left(\left[\frac{\partial F}{\partial \theta}\right]^j\right)\right)(T - F^j)$$

5. Update the initial values:

$$\theta^{j+1} = \theta^j + d\theta^j$$

6. Repeat the process for equations until:
$d\theta^j$ is satisfied with the desired degree of accuracy.
The solution must satisfy the condition

$$0 < \theta_1 < \theta_2 < \cdots < \theta_P < \frac{\pi}{2}$$

## APPENDIX B

MATLAB Program algorithm of a Newton-Raphson algorithm tocalculate switching angles for elimination of third and fifth harmonics.

```
clc; clear all;
A=input('The Modulation index, M=');
M=0:0.01:A;
for ii=1:length(M)
Theta1=35*pi/180;
Theta2=55*pi/180;
Theta3=80*pi/180;
for i=1:100
T=[M(ii)*pi/4 0 0]';
F=[cos(Theta1)-cos(Theta2)+cos(Theta3);
  cos(3*Theta1)-cos(3*Theta2)+cos(3*Theta3);
  cos(5*Theta1)-cos(5*Theta2)+cos(5*Theta3)];
dF=[-sin(Theta1) +sin(Theta2) -sin(Theta3);
   -3*sin(3*Theta1) +3*sin(3*Theta2) -
3*sin(3*Theta3);
   -5*sin(5*Theta1) +5*sin(5*Theta2) -
5*sin(5*Theta3)];
dTheta=(inv(dF))*(T-F);
i
Theta=[Theta1;Theta2;Theta3]*180/pi
F
dTheta*180/pi
Theta1=Theta1+dTheta(1);
Theta2=Theta2+dTheta(2);
Theta3=Theta3+dTheta(3);
if dTheta>-1e-15 & dTheta<1e-15
   break;
end
end
Theta11(ii)=Theta1*180/pi;
Theta22(ii)=Theta2*180/pi;
Theta33(ii)=Theta3*180/pi;
end
plot(M,Theta11,'b','LineWidth',2);
hold on;
plot(M,Theta22,'r','LineWidth',2);
hold on;
plot(M,Theta33,'k','LineWidth',2);
ylim([0 90]);
grid on;
xlabel('Modulation index M');
ylabel('Switching Angles(°)');
title('Switching Angles with N-R Algorithm');
```